\documentclass[sigconf]{acmart}

\usepackage{multicol}
\usepackage{booktabs}
\usepackage{multirow}
\usepackage{siunitx}
\usepackage{graphicx}

\usepackage{subcaption}

\AtBeginDocument{ \providecommand\BibTeX{{ \normalfont B\kern-0.5em{\scshape i\kern-0.25em b}\kern-0.8em\TeX}}}

\copyrightyear{2025}
\acmYear{2025}
\setcopyright{rightsretained}
\acmConference[RecSys '25]{Proceedings of the Nineteenth ACM Conference on Recommender Systems}{September 22--26, 2025}{Prague, Czech Republic}
\acmBooktitle{Proceedings of the Nineteenth ACM Conference on Recommender Systems (RecSys '25), September 22--26, 2025, Prague, Czech Republic}\acmDOI{10.1145/3705328.3748110}
\acmISBN{979-8-4007-1364-4/2025/09}

\begin{document}
\title{Practical Multi-Task Learning for Rare Conversions in Ad Tech}
\author{Yuval Dishi}
\email{yuval.dishi@teads.com}
\affiliation{
  \institution{Teads}
  \country{Israel}
  \city{Netanya}
}
\author{Ophir Friedler}
\email{ophir.friedler@teads.com}
\affiliation{
  \institution{Teads}
  \country{Israel}
  \city{Netanya}
}
\author{Yonatan Karni}
\email{yonatan.karni@teads.com}
\affiliation{
  \institution{Teads}
  \country{Israel}
  \city{Netanya}
}
\author{Natalia Silberstein}
\email{natalia.silberstein@teads.com}
\affiliation{
  \institution{Teads}
  \country{Israel}
  \city{Netanya}
}
\author{Yulia Stolin}
\email{yulia.stolin@teads.com}
\affiliation{
  \institution{Teads}
  \country{Israel}
  \city{Netanya}
}
\begin{abstract}
We present a Multi-Task Learning (MTL) approach for improving predictions for rare (e.g., <1\%) conversion events  in online advertising. 
The conversions are classified into ``rare'' or ``frequent'' types based on historical statistics. The model learns shared representations across all signals while specializing through separate task towers for each type. 
The approach was tested and fully deployed to production, demonstrating consistent improvements in both offline (0.69\% AUC lift) and online KPI performance metric (2\% Cost per Action reduction).

\end{abstract}



\maketitle

\section{Introduction}

In performance advertising \cite{performancebasedadvertising}, the goal is to drive users to take actions that deliver value to advertisers. These actions, called \textit{conversions}, include watching a video, submitting contact details, subscribing to a service, making a purchase, etc. 

The different types of conversions vary widely in frequency and value. For example, purchases of expensive items are much more rare and valuable compared to frequent conversions like landing page views, which are of low~value.

Conversion rate (CVR) prediction models~\cite{rare_events_2010,Response_Prediction_2013,conversion_rate_2012,practical_2017} are a critical component of online advertising systems. They estimate the likelihood that a user will convert after clicking an ad. These predictions are used to rank and price ads in real time, directly impacting marketplace efficiency and advertiser performance.

Predicting \textit{rare} conversions is particularly challenging due to their inherent data sparsity issues, and in particular, the low conversion rate, which can lead to less stable and less accurate predictions.
To address this, we adopt a \textit{multi-task learning} (MTL) approach~\cite{MTL97,ESMM2018,Pan_MTL_conversions_types,MTL,ESCM2,ClickConvSIGIR23,DCMT23}, where rare (referred to as \textit{hard}) and frequent 
 (referred to as \textit{soft}) conversions are modeled as separate tasks. The model learns shared representations from both, using soft conversions as an auxiliary signal to support the hard conversions task.

Multi-task learning has been used to improve CVR prediction in various ways. 
ESMM~\cite{ESMM2018} jointly models CTR and CVR, 
DCMT~\cite{DCMT23} extends this with causal modeling for selection bias, and Wang et al.~\cite{ClickConvSIGIR23} incorporate position bias mitigation. 
The work of Pan et al.~\cite{Pan_MTL_conversions_types} is closest to ours, as they jointly predict different types of conversions to enhance overall accuracy.

Unlike the previous MTL approach, which defines tasks using manually tagged conversion types~\cite{Pan_MTL_conversions_types}, we define tasks based on the historical statistics of the conversion signals, avoiding noisy labels that can degrade model quality. By letting data guide task assignment, we reduce dependency on human tagging and improve robustness at scale.
We deployed the proposed MTL model to production, replacing the previous baseline for hard conversions prediction. This model delivered consistent lifts in
both offline and online performance.
We also addressed serving stability challenges, which are especially important when optimizing low conversion~signal.

Our contributions in this paper are as follows:
\begin{itemize}
  \item A practical MTL framework tailored to hard conversions optimization, using data-driven task definitions rather than manual labeling.
  \item A large-scale deployment of this approach in a live ad system, showed consistent offline and online gains.
  \item A discussion of serving constraints and stability challenges encountered in production.
\end{itemize}

\vspace{-0.2cm}

\section{Background and Motivation}
At Teads\footnote{See https://www.teads.com/the-global-media-platform/}, we operate a dynamic advertising marketplace where content requests from publishers are matched with ad candidates. For each request, we rank and return the best ads based on expected revenue, computed using predictive models for click-through rate (CTR) and post-click conversion rate (CVR). These predictions are used for ranking as well as to calculate bids in our auction~system. 

In this paper, we focus specifically on improving the CVR prediction (pCVR) models.
We use 
Deep \& Cross architecture ~\cite{DeepCross,DeepCross2} implemented in Keras/TensorFlow~\cite{TensorFlow} to estimate pCVR from 
features such as user, context, and ad attributes. 

Historically, we used a single model across all types of conversion events. However, this approach struggles with data imbalance. 
For example, a 
landing page view
(\textit{soft}) event may have a post-click conversion rate 
above 30\%, while a purchase (\textit{hard}) event may be below 0.1\%. 
This wide dynamic range negatively affects model calibration, especially in the low prediction range (e.g., <1\%), where absolute differences are small. Moreover, advertisers with rare but valuable conversions are often high-stakes clients, where even small model improvements can yield significant business impact.

While advertisers generally name and tag their conversion events based on intent, mistakes and inconsistencies can occur. Our approach uses historical outcomes rather than subjective tags to classify data into hard/soft tasks. This ties task assignment to empirical results, ensuring model behavior aligns with observed event frequencies in production.

These observations motivated our task definition 
and the multi-task learning framework (see Section~\ref{sec:Approach}) we present in this~work.

\section{Our MTL Approach}
\label{sec:Approach}
\subsection{Offline Training}
    \subsubsection{Model Architecture}
\begin{figure}[h!]
            \centering
            \includegraphics[width=0.85\columnwidth]{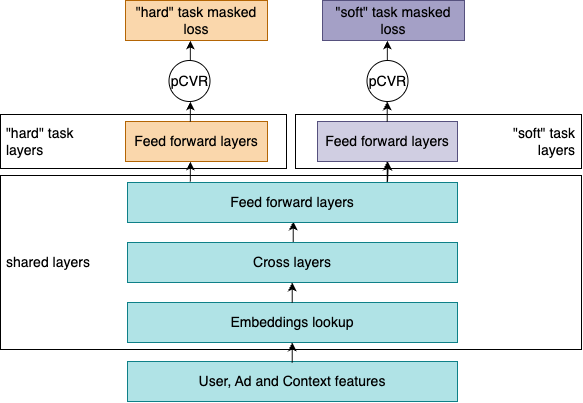}
            \caption{MTL Architecture}
            \label{fig:mtl_arch2}
            \vspace{-0.5cm}
            \end{figure}

  Our multi-task learning (MTL) model shares a common embedding layer and, optionally, additional cross and feed-forward layers, as shown in Figure~\ref{fig:mtl_arch2}. The network then splits into two task-specific towers, one for soft conversions and one for hard conversions. These towers receive the same input and have identical architecture. 
  We apply minibatch training, and each example contributes only to the loss component of its assigned task. The overall minibatch loss is computed as:  
\begin{equation}
    \label{eq: loss}
\mathcal{L} = W_{\text{soft}} \cdot \mathcal{L}_{\text{soft}} + W_{\text{hard}} \cdot \mathcal{L}_{\text{hard}}.
    \end{equation}
We control the number of shared vs. task-specific layers, and treat both this setting and task loss weights $W_{\text{soft}}$, $W_{\text{hard}}$ as hyperparameters. 
This setup enables learning shared representations while focusing optimization on the more challenging hard conversion task. Hyperparameters are tuned via offline research (see Section~4.1).
\subsubsection{ Definition of hard/soft tasks}
\label{sec:def-hard-soft}
    In our MTL setup, we define two tasks based on the historical statistical CVR of each conversion setup: soft conversions with CVR > $\alpha$ and hard conversions with CVR~$\leq\alpha$. The threshold $\alpha$ is selected via offline validation to align with typical patterns, such as purchases and certain advertiser types tend to fall into the hard category. Historical CVR is computed as a time-decayed ratio of conversions to clicks over a recent sliding window, giving more weight to recent data. This data-driven segmentation avoids relying on potentially inaccurate or inconsistent metadata like conversion category or advertiser~type.

\subsection{Online Setting}
\subsubsection{Serving Setup}
In the serving phase, we adopted a concept of an inference model, where only the relevant task-specific branch (in our case, the hard-conversions) is retained from the full model graph, effectively resulting in a single prediction model. 
Although training involves all tasks, inference remains as efficient as with the baseline model, since unrelated components are removed. This setup retains the generalization benefits of shared training without added serving cost.
\vspace{-6pt}
 \subsubsection{Traffic Routing}
For the serving of our MTL model in general and for an online A/B test in particular, we built a component that classifies each ad candidate as soft or hard and routes it to the corresponding model. The traffic splitter operates within the variant logic and uses a threshold-based decision on the candidate’s classification. To improve stability for borderline cases, the online threshold is set slightly lower than the one used offline (see Section~\ref{sec:def-hard-soft}).
\vspace{-6pt}
\subsubsection{Stability Challenges}
A potential challenge in categorizing conversions based on historical CVR arises from its inherent instability. Performance fluctuations can cause a conversion to be classified as ``soft'' one day and ``hard'' the next, leading to inconsistencies in training and serving. To mitigate this, we developed a stabilization mechanism that incorporates multiple contextual elements, such as the historical CVR over a predefined period, the advertiser type, the conversion category, and other relevant data.
\section{Results}
\subsection{Offline Evaluation}
\label{sec:offline}
\subsubsection{Offline Metrics}   
For offline evaluation, we run experiments on hundreds of millions of logged Teads records, comparing each model’s performance using Relative Information Gain (RIG)~\cite{metrics},
a linear transformation of log-loss given by:
\begin{equation}
RIG = 1-\frac{-c\cdot log(p)-(1-c)log(1-p)}{-\gamma\cdot log(\gamma)-(1-\gamma)log(1-\gamma)},
\end{equation}
where $c$ is the observed conversion, $p$ is the predicted probability, and  $\gamma$ is the average CVR in the evaluation set.
We also report Area Under the Curve (AUC)~\cite{hanley1982auc} in the main comparison between MTL and the baseline model\footnote{AUC was omitted from other comparisons due to the high computational cost on large datasets.}.

\subsubsection{Offline Results}
In the offline phase, we focused on optimizing the model performance for hard conversions through systematic hyperparameter tuning of the MTL setup. We first tuned the task-specific loss weights $W_{\text{hard}}$ and $W_{\text{soft}}$ (see Eq.\eqref{eq: loss}), comparing multiple configurations. Results are shown in Table~\ref{tab:RIG_results}, where all values are reported relative to a baseline model with $W_{\text{hard}}=1.0, W_{\text{soft}}=0$  which is equivalent to training on hard conversions only. The best performance was achieved with balanced weights ($W_{\text{hard}}=W_{\text{soft}}=0.5$), demonstrating that using soft conversions as an auxiliary signal helps improve generalization on hard conversions.

After selecting the best loss weights, we tuned the number of shared layers.
We evaluated several values and compared model performance to a baseline with no shared layers (sharing only embeddings). As shown in Table~\ref{tab:RIG_results2}, sharing 
more layers led to further gains, supporting the benefit of shared representation learning.

\begin{table}[ht]
  \centering
  \begin{minipage}[t]{0.48\linewidth}
    \centering
    \vspace{0pt}
    \begin{tabular}{lcc}
      \hline
      \textbf{$W_{\text{hard}}$} & \textbf{$W_{\text{soft}}$} & \textbf{RIG} \\
      \hline
      0.2 & 0.8 & +0.86\% \\
      0.5 & 0.5 & +1.17\% \\
      0.6 & 0.4 & +1.16\% \\
      0.7 & 0.3 & +1.04\% \\
      0.8 & 0.2 & +0.8\%  \\
      0.9 & 0.1 & +0.55\% \\
      \hline
    \end{tabular}
    \caption{
Loss Weights} \label{tab:RIG_results}
  \end{minipage}
  \hfill
  \begin{minipage}[t]{0.48\linewidth}
    \centering
    \vspace{0pt}
    \begin{tabular}{lc}
      \hline
      \textbf{Shared } & \textbf{RIG} \\
      \textbf{Layers} &  \\
      \hline
      1 & +0.03\% \\
      2 & +0.11\% \\
      3 & +0.19\% \\
      4 & +0.21\% \\
      \hline
    \end{tabular}
    \caption{
    Number of Shared Layers}
     \label{tab:RIG_results2}
  \end{minipage}
\end{table}

\vspace{-20pt}

Finally, we compared our best MTL model to the production baseline, a single-task model trained on all conversion types. The comparison was conducted on the hard conversions population, which is our primary focus. Results, summarized in Table~\ref{tab:offline_results}, show that the MTL approach outperforms the baseline, validating its effectiveness for predicting 
rare but high-value conversions.
\vspace{-3pt}
\begin{table}[H]
\centering
\begin{tabular}{lcc}
\hline
\textbf{Method} & \textbf{RIG}  & \textbf{AUC} \\
\hline
Best MTL (with tuned hyperparameters) & +4.08\% & +0.69\% \\
\hline
\end{tabular}
\caption{MTL vs One Task Baseline CVR Model}
\label{tab:offline_results}
\end{table}
\vspace{-0.9cm}
\subsection{Online Evaluation}
In this section, we present the online evaluation of the MTL model, selected as the best-performing model from our offline experiments (see Section 4.1), in comparison to the baseline (one-task) model currently deployed in production.
\vspace{-6pt}
\subsubsection{A/B Test Setup}   
The MTL model was trained using several months of historical log data. 
For the online evaluation, the model was deployed in our A/B testing framework and served to a subset of live production traffic on the Teads platform. 

We conducted the A/B test using a budget split setup, where the control and variant received equal share of traffic and equal portion of each advertiser's budget. This design ensures a fair comparison by preventing spend leakage and eliminating biases from uneven auction dynamics.
\subsubsection{Online KPIs}
For online evaluation of a given conversion setup, we use two key metrics.
\begin{itemize}
    \item \textit{CPA ratio}: The ratio of the Cost per Action (CPA) under the variant (MTL model) to the CPA under the control (baseline model). A CPA ratio < 1 indicates improved cost efficiency.
    \item \textit{Conversion ratio}: The ratio of the number of conversions in the variant to the number in the control. A ratio > 1 indicates improved conversion volume.
\end{itemize}

To ensure a fair and comprehensive assessment, we compute the
CPA and Conversion ratios for each conversion setup individually, and then apply several aggregation methods to capture both fairness and business impact.
\begin{itemize}
    \item \textit{Median:} Treats all advertisers equally, computing the median CPA and Conversion ratios across all.
    \item \textit{Spend-Weighted Geometric Mean:} Weighs each advertiser’s metric by their spend, emphasizing performance for high-investment advertisers. The spend-weighted geometric mean is computed as 
    $\text{exp}\left( \frac{\sum_{i=1}^{N} s_i \cdot \text{ln} (m_i)}{\sum_{j=1}^{N} s_j}\right)$,
   where $s_i$ is the spend and $m_i$ is the ratio (CPA or Conversion) for advertiser $i$.
   \item \textit{High-Impact Advertiser Segment:} To assess performance on strategically important accounts, we compute CPA and conversion ratios over a segment of high-impact advertisers, selected based on their business significance and activity~scale.
\end{itemize}
These KPIs together provide a comprehensive view of model impact, balancing advertiser fairness, strategic value, and overall efficiency.
\vspace{-15pt}
\subsubsection{Online Results}
    We evaluated the MTL model against the production baseline model in an A/B test, focusing on the hard conversions segment, which is our key optimization target.

Table~\ref{tab: online_results} summarizes the results across all aggregation methods, consistently showing that the MTL model achieves a lower CPA ratio, indicating reduced cost per action, and a higher Conversion ratio, reflecting an increase in conversion volume.
Specifically, 
\begin{itemize}
    \item Median aggregation shows broad improvement across the advertisers.
    \item Spend-weighted metrics confirm that the gains are preserved (or amplified) among high-spend advertisers.
   \item We define three nested groups of advertisers (\textit{Group~I},\textit{ Group~II}, and \textit{Group~III}), representing 
   subsets of high-impact advertisers, contributing approximately 30\%, 50\%, and 70\% of overall platform activity, respectively.
\end{itemize}
These results demonstrate that the MTL model achieves better performance, both in aggregate and for our key advertisers.
\vspace{-7pt}
            \begin{table}[H]
            \centering
            \begin{tabular}{lcc}
            \hline
            \textbf{Metric Type} & \textbf{CPA Ratio} & \textbf{Conversion ratio} \\
            \hline
            Median & -2.0\%  & +2.6\% \\
            Spend-Weighted & -3.4\% & +3.0\%  \\
            Group I & -5.1\% & +4.8\%  \\
            Group II & -4.1\% & +4.8\%  \\
            Group III & -2.4\% & +2.2\%  \\
            \hline
            \end{tabular}
            \caption{Online Results}
            \label{tab: online_results}
            \end{table}
\vspace{-0.9cm}
\section{Conclusion}
We introduced a multi-task learning approach to improve prediction of rare, high-value conversion events by leveraging shared representations learned from more frequent signals. Tasks are automatically defined using historical conversion statistics, eliminating the dependency on advertisers’ tagging. 
The described approach has become part of our production system and, along with other model improvements, powers all hard-conversion predictions across our platform. 
This work may also apply to other rare event prediction tasks in large-scale advertising systems, such as low CTR events or other infrequent user actions.



\section*{Speaker Bio}

\textbf{Yuval Dishi} is a senior machine learning engineer and a tech lead in the AI department at Teads where he's responsible for enhancing and developing algorithms for a variety of recommendation system components.
Before joining Teads, he worked on research projects in the fields of ML and cyber security.
He holds a MSc from the Computer Science Department at the Bar-Ilan University.
Research interests include deep learning, classical machine learning, and crowdsourcing. 

\textbf{Dr. Natalia Silberstein} is a Data Science Team Lead in the AI department at Teads (formerly Outbrain), where she focuses on developing and improving machine learning models and algorithms for personalized ad selection. Prior to joining Outbrain, she worked as a research scientist at Yahoo Research Haifa. She holds a Ph.D. from the Department of Computer Science at the Technion – Israel Institute of Technology. Following her doctoral studies, she completed postdoctoral research in the Department of Electrical and Computer Engineering at the University of Texas at Austin.



\bibliographystyle{ACM-Reference-Format}
\bibliography{main}


\end{document}